\documentclass[preprint,amsmath,amssymb,11pt,aps,showpacs,showkeys,pra]{revtex4-2}
\usepackage{graphicx}
\usepackage{graphics}
\usepackage{amsmath}
\usepackage{dcolumn}
\usepackage{amssymb}
\usepackage{bm}
\usepackage[utf8]{inputenc}

\begin{document}
\title{Redshift as a stretching factor in rotating graphene wormholes}
\author{Everton Cavalcante}\email{Corresponding Author: everton@servidor.uepb.edu.br}
\affiliation{Departamento de Física, Universidade Estadual da Para\'{\i}ba, Campina Grande, PB, Brazil - Corresponding Author: everton@servidor.uepb.edu.br}
\author{Claudio Furtado}\email{furtado@fisica.ufpb.br}
\affiliation{Departamento de Física,  Universidade Federal da Paraíba, João Pessoa, PB, Brazil}


\begin{abstract}

In this paper, we discuss an extension of the geometric description of graphene wormholes in a non-inertial situation. We present an effective metric that describes the wormhole connection between two graphene sheets with matter content in rotation. Additionally, a stretching term as a function of the classical redshift of space has been found and discussed. We also explore the influence of a rotation term on quantum holonomy, recovering previous results found for the static case.

\end{abstract}



\maketitle

\section{Introduction}\label{secI}

General relativity (GR) is one of the most profound theories in modern physics. Albert Einstein's formulation of GR in 1915 replaced Newtonian gravity and provided a new way to understand the force that governs the motion of celestial bodies. One of the most remarkable predictions of GR is the phenomenon known as the redshift of light. This effect has far-reaching consequences, especially in the context of cosmology \cite{DodelsonSchmidt}. At the heart of general relativity are Einstein's field equations, a set of equations that describe the relationship between the curvature of spacetime and the distribution of matter and energy within it. These equations can be expressed in a simplified form as $G_{\mu\nu} = 8\pi G T_{\mu\nu}$, where \(G_{\mu\nu}\) represents the Einstein tensor, a mathematical object that encodes the curvature of spacetime, and \(T_{\mu\nu}\) is the energy-momentum tensor, which describes the distribution of matter and energy in the universe. \(G\) is the gravitational constant.

The set of GR equations is widely acknowledged for its prediction that massive celestial objects, such as stars or galaxies, generate gravitational fields capable of warping the fabric of spacetime in their vicinity. As particles traverse through this curved spacetime, their paths become elongated, resulting in energy loss. Consequently, there is a change in their wavelength, leading to the phenomenon known as gravitational redshift. The mathematical expression for the redshift factor (\(z\)) is given by 
\begin{equation}
z = \frac{\lambda_{\text{observed}} - \lambda_{\text{emitted}}}{\lambda_{\text{emitted}}} \mbox{.}
\end{equation}
where \(\lambda_{\text{observed}}\) is the observed wavelength of light, and \(\lambda_{\text{emitted}}\) is the wavelength of light when it was emitted.

The initial experimental validation of gravitational evidence emerged through observations of the Sun. In 1919, Sir Arthur Eddington led an expedition to the island of Príncipe, and the other to the Brazilian town of Sobral, to witness a total solar eclipse. During the eclipse, Eddington measured the apparent positions of stars close to the limb of the Sun and compared them to their positions when the Sun was not in proximity. His findings revealed that the starlight was bent as it passed near the Sun, aligning with Einstein's predictions. This marked the inaugural observational confirmation of general relativity (GR) and gravitational evidence \cite{DysonEddingtonDavidson}.

In the cosmological framework based on GR, redshift primarily arises from the expansion of space. This implies that the farther a galaxy is from us, the more space has expanded since the light left that galaxy. Consequently, the light becomes more stretched and exhibits a higher degree of redshift. Hubble’s law, coupled with the concept of the redshift factor, provided strong observational support for the expanding universe and led to the development of the Big Bang theory.

Beyond the scene of the traditional general relativity (GR) scenario within cosmology, the scientific community has been closely monitoring the advancements in low-dimensional physics for several decades. There is a growing interest in exploring the applicability of the GR framework to these systems \cite{Boozer, JPSLemos}. Notably, graphene stands out as one of the most prominent low-dimensional materials in contemporary research. The convergence of graphene and GR has been substantiated by some influential papers, including \cite{KatanaevandVolovich, FialkovskyVassilevich, KleinertGaugeFieldsinCondensedMatter}.  These papers support the thesis that it is possible to describe defects in solids through a differential geometry approach. So these defects now assume the function of mass in traditional cosmology. The main advantage of drawing this way is the powerful machinery of GR.
With respect to such an approach for addressing defects in the graphene lattice—now referred to as topological defects—these can be replicated using a non-Abelian gauge field source. This bestows upon graphene the distinction of serving as a handheld laboratory for a 2+1-dimensional quantum field theory (QFT) within the context of a topological defect density background.

The convergence of quantum field theory in curved spacetimes with graphene physics reveals an intriguing theoretical domain, where the realms of quantum mechanics and general relativity harmoniously intersect within a condensed matter system. The behavior of low-energy electron excitations in graphene, governed by a massless Dirac equation, strikingly mirrors the characteristics of relativistic particles navigating a curved space-time. This inherent correspondence provides  physicists with a unique opportunity to investigate analogs of gravitational effects on quantum fields within the confines of a condensed matter environment.

In a preceding article authored by one of us, we delved into the concept of a graphene wormhole, formed by the junction of two graphene sheets connected by a nanotube, within the framework of Quantum Field Theory (QFT). This work \cite{EvertonWormhole}, involved the derivation and analysis of quantum holonomies. In the GR scenario, the first solution for wormholes was put forward by Einstein \cite{EinsteinRosen} as a possible bridge that was made possible by a coordinate change in Schwarzchild geometry. Here it is important to emphasize that, unlike the cosmological solution, wormholes in graphene do not need to satisfy the energy conditions described in \cite{MorrisThorne}.  These varieties of wormholes are built by two graphene sheets junction with a nanotube  within a zigzag boundary connection on the bridge. It is also been assumed a large radius of the wormhole hole. Bigger than the throat length. This way we avoid the effects of soft variations on the junctions of the bridge. 
Actually, recently some papers have shown the radius of the throat must be closer than $70$ Angstrons \cite{ElllisBronikovRamosFurtado, catenoidbridge}, which represents around 50 times the usual interatomic distance in the lattice ($d \approx 1,42$ Angstrons) \cite{Katsnelson}.

In this paper, our focus is on revisiting the issue outlined in \cite{EvertonWormhole}, however now incorporating the matter content in a non-inertial situation. We maintain the inclusion of the proposed redshift term, as indicated in \cite{MorrisThorne, ElllisBronikovRamosFurtado}, ignored in previous research due to the static nature of the problem. In Sect. 2, we obtain that the redshift term can be seen as an interaction of the lattice gauge with some stretching field acting on the material. Acting as a new approach particularly applicable to scenarios involving material subjected to mechanical stress. Additionally, our analysis reveals the derivation of constraints on the azimuthal solutions of the Dirac equation for massless fermions.

Moving on to Section 3, we extend our investigation to evaluate the impact of rotation factors on the holonomy matrices. This examination provides insights into the rotational terms on the geometric phases associated with parallel transport around closed loops in the material. Finally, in Section 4, we present our conclusions drawn from the study and discuss potential applications.

\section{Graphene wormhole geometry in a non-inertial frame}\label{secII}

The metric around a traversable wormhole within the Morris and Thorne framework can be written (in terms of the redshift $\phi(r)$) generally as:

\begin{equation}\label{metrica geral}
ds^2=-e^{2\phi(r)}dt^2 + \frac{dr^2}{1-\frac{b(r)}{r}} + r^2 d\theta^2 + r^2 \sin^2 \theta d\phi^2 \mbox{.}
\end{equation}

For a moment, it is appropriate to discuss that one of the first problems wormholes bring out is their instability in an electromagnetic context \cite{MislerWheeler}. Such instability can be avoided if the metric has some preconditions. Such as throat and flare-out requirements \cite{ElllisBronikovRamosFurtado, Ellis, Bronnikov, MorrisThorne2}.
This stability, and the possibility for fermions to pass through the throat, can be achieved when the flare-out
$\frac{r}{b(r)} -1> 0$, and throat conditions: $\frac{b(r)-\dot{b}(r)r}{2b(r)^2} >0$  are satisfied for a  metric of a general wormhole \cite{NonviolationconditionsGodani, StabilitythinshellGodani}. 
Both of them already achieved for the metric in (\ref{metrica geral}).

Within the context of the wormhole in graphene, we have to make some adaptations to the metric described above. The first is to assume the speed of the fermions inside the structure as Fermi ($v_{f}$). We must to adapt too the model to the fact that the regions inside and outside the wormhole are covered by different metrics. Each one adapted to the geometry of each region. As well exemplified in previous works \cite{EvertonWormhole, ElllisBronikovRamosFurtado, GonzalezHerrero, GarciaFurtado}. We also assume the shape function is negligible ($b(r) \ll r$), and that the model is based on a symmetry of the type: $d\phi \to 0$. Assuming further that the function is regular everywhere, we can introduce the Heaviside step function ($\Theta(r)$) through a conformal factor $\Omega(r)$ as:

\begin{equation}\label{metrica p wormhole}
ds^2=\Omega^2(r) \bigg [
-v_{f}^2 e^{2\phi(r)}dt^2 + dr^2 + r^2 d\theta^2 \bigg ]
\end{equation}

This way we can obtain a specific metric for both regions: inside ($r \le R$) and outside ($r \ge R$) the throat. Where the Heaviside function and the metric relate as:

\begin{eqnarray}
g_{\mu \nu}(x)= \Omega^2(r)
\left(\begin{array}{cccc}
- \tilde{v}_{f}^2 & 0 & r^2 \omega_{\theta} \\ 
0 & 1 & 0  \\
r^2 \omega_{\theta} & 0 & r^2 \\
\end{array}\right) \mbox{,}
\qquad
\Omega (r)= \bigg ( \frac{R}{r} \bigg )^2 \Theta (R-r)+\Theta (r-R) \mbox{.}
\end{eqnarray}



The rotation of the molecule structure is introduced in a manner analogous to what was done in reference \cite{EvertonClaudio2}. Here we do a simple coordinate transformation: \( \theta \to \theta^{'}=\theta +\omega_{\theta}t \), where \( \omega_{\theta} \) is the angular velocity of the rotating frame. 
Here it is interesting to point out that the term \( \omega_{\theta} \) must to obey some limits to harmonize with the principles of relativity. As the speed of the fermions in our model is around the Fermi speed ($v_{f}$), we will assume here that the product of the \( \omega_{\theta} \) term with the distance to the singularity must be smaller than the Fermi speed in the lattice (\( \omega_{\theta} r < v_{f} \)).

The new line element is then rewritten as:

\begin{equation}\label{métrica com rotação}
ds^2=\Omega^2(r) \bigg [
-v_{f}^{2}e^{2\phi(r)}dt^2 + dr^2 + r^2 (d\theta + \omega_{\theta}dt)^2  \bigg ]
\end{equation}
For facility, we assume: $\tilde{v}_{f}^2=v_{f}^2 e^{2\phi(r)} - r^2 \omega_{\theta}^2 $, and the line element become as follows:

\begin{equation}\label{métrica com rotação 2}
ds^2=\Omega^2(r) \bigg [
-\tilde{v}_{f}^{2}dt^2 + dr^2 + r^2 d\theta + 2 \omega_{\theta} r^2 d\theta dt  \bigg ]
\end{equation}

The geometric structure of this  curved space  are described by local reference bases  known as tetrads (${e^{a}}_{\mu}(x)$), which are defined at each point in space-time by a local reference
frame $g_{\mu \nu}(x)=\eta_{ab}{e^{a}}_{\mu}{e^{b}}_{\nu}$. The tetrads and its inverse, ($e^{\mu}_{a}=\eta_{ab}g^{\mu \nu}e^{b}_{\nu}$), 
satisfy the orthogonal relationships: $e^{a}_{\mu}e^{b \mu}=\eta^{ab}$, $e^{a}_{\mu}e^{\mu}_{b}=\delta^{a}_{b}$, $e^{\mu}_{a}e^{a}_{\nu}=\delta^{\mu}_{\nu}$, 
and map the curved  reference frame via the local reference frame \cite{Birrel}:
\begin{equation}
ds^{2}=g_{\mu \nu}dx^{\mu}dx^{\nu}=e^{a}_{\mu}e^{b}_{\nu}\eta_{ab}dx^{\mu}dx^{\nu}=\eta_{ab}\theta^{a}\theta^{b} \mbox{.}
\end{equation}
Greek indices ($\mu, \nu$) run for space-time frame coordinates and the Latin indices ($a,b$) run for local frame coordinates. 
For the local reference frame we must define a new basis as: ($\theta^{a}=e^{a}_{\mu}(x)dx^{\mu}$). Now we write the set of tetrads for this geometry. The tetrad and its inverse are defined as: 

\begin{eqnarray}
{e^{a}}_{\mu}(x)=\frac{\Omega(r)}{\sqrt{2}}
\left(\begin{array}{cccc}
i(r\omega_{\theta} - iv_{f}e^{\phi}) & 0 & ir \\ 
0 & \sqrt{2} & 0  \\
r\omega_{\theta} + iv_{f}e^{\phi} & 0 & r \\
\end{array}\right) \mbox{,} 
\quad
{e^{\mu}}_{a}(x)= \frac{{\Omega}^{-1}(r)}{\sqrt{2}v_{f}e^{\phi}}
\left(\begin{array}{cccc}
1 & 0 & -i \\ 
0 & \sqrt{2}v_{f}e^{\phi} & 0  \\
-\frac{1}{r}(r\omega_{\theta} + iv_{f}e^{\phi}) & 0 & \frac{i}{r}(r\omega_{\theta} - iv_{f}e^{\phi}) \\
\end{array}\right)
\end{eqnarray}





The beautiful way to obtain the one-form connection for QFT in curved spaces is based on the first  Maurer-Cartan structure equations \cite{Birrel}:
\begin{equation}
d\theta^{a}+{\omega^{a}}_{b} \wedge \theta^{b}=0 \mbox{.}
\label{Maurer-Cartan equation}
\end{equation}
To find the non-null components of 1-form connections we must remind with $d(r^{-1})=\frac{\partial }{\partial r}r^{-1}dr=-\frac{1}{r^2}dr$, and $d(r^{-2})=\frac{\partial }{\partial r}r^{-2}dr=-\frac{2}{r^3}dr$. We also name the term $S_{sf}=\frac{v_{f}e^{\phi}}{r}$ as a "stretching factor term". Thus, the 1-form connections are given by the following expressions:

\begin{eqnarray}\label{conexões de 1-forma dentro da garganta}
\begin{cases} 
{{\omega_{t}}^{0}}_{1}=-{{\omega_{t}}^{1}}_{0}= - \bigg ( \frac{i}{\sqrt{2}}\omega_{\theta} -\sqrt{2} S_{sf} \bigg ), \\
{{\omega_{t}}^{2}}_{1}=-{{\omega_{t}}^{1}}_{2}= - \bigg ( \frac{1}{\sqrt{2}}\omega_{\theta} + i \sqrt{2} S_{sf} \bigg ), \\
{{\omega_{\theta}}^{0}}_{1}=-{{\omega_{\theta}}^{1}}_{0}=-\frac{i}{\sqrt{2}}, \\
{{\omega_{\theta}}^{2}}_{1}=-{{\omega_{\theta}}^{1}}_{2}=-\frac{1}{\sqrt{2}},
 \end{cases}
\mbox{;} \qquad \textrm{for} \quad r \le R \mbox{,}
\end{eqnarray}
and
\begin{eqnarray}\label{conexões de 1-forma fora da garganta}
\begin{cases} 
{{\omega_{t}}^{0}}_{1}=-{{\omega_{t}}^{1}}_{0}=\frac{i}{\sqrt{2}}\omega_{\theta}, \\
{{\omega_{t}}^{2}}_{1}=-{{\omega_{t}}^{1}}_{2}=\frac{1}{\sqrt{2}}\omega_{\theta}, \\
{{\omega_{\theta}}^{0}}_{1}=-{{\omega_{\theta}}^{1}}_{0}=\frac{i}{\sqrt{2}}, \\
{{\omega_{\theta}}^{2}}_{1}=-{{\omega_{\theta}}^{1}}_{2}=\frac{1}{\sqrt{2}},
 \end{cases}
\mbox{;} \qquad \textrm{for} \quad r \ge R \mbox{.}
\end{eqnarray}

Using the first equation of structure of Cartan we obtain the following non-null components  of spinorial connections  $\Gamma_{\mu}(x)=\frac{i}{4}\omega_{\mu ab}\Sigma^{ab}$,  that are given by

\begin{eqnarray}
\begin{cases}
\Gamma_{t}(x)=\frac{i}{2\sqrt{2}}\bigg ( \omega_{\theta} + 2i S_{sf} \bigg )\sigma^3 \\
\Gamma_{\theta}(x)=\frac{i}{2\sqrt{2}}\sigma^3 \\
\end{cases}
\mbox{,}  \quad \textrm{for} \quad r \le R \mbox{;}
\qquad
\begin{cases} 
\Gamma_{t}(x)=-\frac{i}{2\sqrt{2}}\omega_{\theta}\sigma^3 \\
\Gamma_{\theta}(x)=-\frac{i}{2\sqrt{2}}\sigma^3 \\
\end{cases}
\mbox{,} \quad \textrm{for} \quad r \ge R \mbox{.}
\end{eqnarray}

Remember that all arrangements constructed differently from the typical (six vertices) in graphene sheets are referred to as topological defects.
\cite{Katsnelson}. And more, each of these defects has associated a fictitious gauge flux in the continuous regime \cite{FialkovskyVassilevich}. So, this work intends to describe the massless fermions by a Weyl equation in the background of a gauge field ($A_{\mu}$) bring about by the presence of heptagon rings in the wormhole nanotube junctions:  
\begin{equation}
i\hbar v_f {e^{\mu}}_{a}\sigma^{a} \big ( \partial_\mu + \Gamma_{\mu}(x) - i A_\mu \big ) \psi =0.
\label{Weyl equation}
\end{equation}

Two further points to consider would be that the, in QFT in curved spaces, the $\vec{\sigma}$ matrices must be rewritten in a new basis as $\sigma^{\mu}={e^{\mu}}_{a}\sigma^{a}$. Or yet, considering $\tilde{\sigma}=\sigma^0 + i\sigma^2 $, and ${\tilde{\sigma}}^{\dagger}=\sigma^0 - i\sigma^2 $, they are made more clearly as:

\begin{eqnarray}
\begin{cases}
\sigma^t=\frac{1}{\sqrt{2}v_{f}e^{\phi}}{\Omega}^{-1}(r) \tilde{\sigma}^{\dagger} \\
\sigma^r={\Omega}^{-1}(r) \sigma^1 \\
\sigma^{\theta}=\frac{1}{\sqrt{2}  v_{f}e^{\phi}}
r^{-1}{\Omega}^{-1}(r)
\big ( -r\omega_{\theta} \tilde{\sigma}^{\dagger} - iv_{f}e^{\phi} \tilde{\sigma}\big ) \\
\end{cases}
\end{eqnarray}
Also, the gauge field ($A_{\mu}$) can be written by a flux ($\Phi$) from the global contribution of each heptagon ring on the wormhole junction. So, the gauge field follows as: 
\begin{equation}
A_\theta = \pm \frac{\Phi}{2\pi}.
\label{flux of global contribution}
\end{equation}

Thus, the Dirac massless equation for both regions of the wormhole throat is given by:

\begin{equation}\label{Weyl2}
 \frac{i\hbar \Omega^{-1}(r)}{\sqrt{2}e^{\phi}} \bigg [
 {\tilde{\sigma}}^{\dagger} \big ( \partial_{t} + \Gamma_{t} \big ) + \sigma^1\partial_{r} - 
\big ( \omega_{\theta} \tilde{\sigma}^{\dagger} + i S_{sf}\tilde{\sigma} \big ) \big ( \partial_{\theta} + \Gamma_{\theta} -i A_{\theta} \big )
\bigg ]  \psi = 0.
\end{equation}

To make the chart of both regions clear, it is convenient to assume: $\beta=\omega_{\theta}+iS_{sf}$ and $\beta^{*}=\omega_{\theta}-iS_{sf}$. So we have:

\begin{equation}\label{Weyl inside 1}
 \bigg [ \frac{\sigma^0 r^2}{R^2} \big ( \partial_{t} - \beta \partial_{\theta} + i\beta A_{\theta} \big ) + \frac{\sigma^1 r^2}{R^2} \bigg ( \partial_{r} +\frac{i \omega_{\theta}}{2\sqrt{2}}  + \frac{i}{2\sqrt{2}} \beta \bigg )
 -\frac{i \sigma^2 r^2}{R^2} \big ( \partial_{t} - \beta^{*} \partial_{\theta} + i \beta^{*} A_{\theta} \big ) 
 +\frac{r^2}{R^2}\frac{S_{sf}}{2\sqrt{2}}\sigma^3
 \bigg ] \psi = 0
 \mbox{,} \quad \textrm{for} \quad r \le R \mbox{,}
\end{equation}
and
\begin{equation}\label{Weyl outside 1}
\bigg [ \sigma^0 \big ( \partial_{t} - \beta \partial_{\theta} + i\beta A_{\theta} \big ) + \sigma^1 \bigg ( \partial_{r} -\frac{i \omega_{\theta}}{2\sqrt{2}}  - \frac{i}{2\sqrt{2}} \beta^{*} \bigg ) 
 -i \sigma^2 \big ( \partial_{t} + \beta^{*} \partial_{\theta} - i \beta^{*} A_{\theta} \big ) 
 +\frac{S_{sf}}{2\sqrt{2}}\sigma^3
 \bigg ] \psi = 0
\mbox{,} \quad \textrm{for} \quad r \ge R \mbox{.}
\end{equation}

Futhermore, the flux in (\ref{flux of global contribution}) can be indexed by region of the wormhole throat. $\tilde{\Phi}$ and $\bar{\Phi}$ for the regions inside and outside, respectively. For that we must to consider ($\tilde{\Phi}=-\bar{\Phi}$) to obtain well-behaved solutions over the entire background geometry.
Another concern might be that how bring up the Fermi points and the sub-lattice index. Particularly the following ansatz has been chosen:
\begin{equation} \label{ansatz}
\psi=e^{\frac{i Et}{\hbar}}\left(\begin{array}{cccc}\Psi^{\nu}_{A}  \\ \Psi^{\nu}_{B} \end{array}\right) = 
e^{\frac{i Et}{\hbar}} \psi^{\nu}_{A,B}
, \qquad \textrm{where} \quad \nu=\pm 1.
\end{equation}
Using that, the equations (\ref{Weyl inside 1}) and (\ref{Weyl outside 1}) for massless fermions results as: 

\begin{equation}\label{Weyl inside 2} 
\Bigg ( \frac{\sigma^{1}r^{2}}{R^2} \bigg ( \partial_{r} + \frac{i\omega_{\theta}}{\sqrt{2}}  - \frac{S_{sf}}{2\sqrt{2}} \bigg ) - \big ( \beta \sigma^0 + i\beta^{*}\sigma^2  \big ) \frac{r^2}{R^2} \bigg ( \partial_{\theta} +  \frac{i\nu }{2\pi} \tilde{\Phi} \bigg ) + \frac{r^2}{R^2}  \frac{S_{sf}}{2\sqrt{2}}\sigma^3 \Bigg ) \psi^{\nu}_{A,B} = 0
\mbox{,} \quad \textrm{for} \quad r \le R \mbox{,}
\end{equation}
and
\begin{equation}\label{Weyl outside 2}
\Bigg ( \sigma^1 \bigg ( \partial_{r} - \frac{i\omega_{\theta}}{\sqrt{2}}  + \frac{S_{sf}}{2\sqrt{2}} \bigg )
- \big ( \beta \sigma^0 + i\beta^{*}\sigma^2  \big ) \bigg ( \partial_{\theta}-  \frac{i\nu }{2\pi} \bar{\Phi} \bigg ) + \frac{S_{sf}}{2\sqrt{2}}\sigma^3 \Bigg ) \psi^{\nu}_{A,B} = 0
\mbox{,} \quad \textrm{for} \quad r \ge R \mbox{.}
 \end{equation}

As is evident in the equation above, the stretching factor remains as gauge coupling in the Dirac equation. Which brings some intriguing consequences to the model. We widely know when subjected to stress, graphene exhibits remarkable mechanical strength. The introduction of strain in graphene can lead to intriguing electronic effects, altering its band structure and electronic transport properties. Understanding the response of graphene to mechanical stress is crucial for engineering applications, such as strain-tunable devices \cite{Straintunablebilayer, TunablegaugefieldWeylsemimetals}. Our proposal here is to interpret this new coupling as an interaction of the network with an external stress field. Which is distributed more intensely in the vicinity of the hole's throat ($S_{sf} \approx \frac{1}{r}$). And, as soon as $S_{sf} \sim \sigma^3$, not change the sublattices too. This new result still requires a more precise discussion in experimental terms. However, it raises the possibility of theoretical manipulation, to say the least, intriguing.

 At this point, it seems appropriate to note that, in the limit $\omega_{\theta} \to 0$, and $v_{f}e^{\phi} \to 1$, we found: $\beta \to i/r$, $ \beta^{*} \to -i/r$, and $S_{sf} \to 1/r$. Where we recovered the equations of the static case \cite{EvertonWormhole}, despite the terms related with $\sigma^0$ and $\sigma^3$. The fact that these terms are "left over" makes us obtain a new result. We can establish a link between the azimuthal solution of the eigenstates, such as:

\begin{equation}\label{vinculo solução azimutal}
\begin{cases} \frac{\partial}{\partial \theta}\psi^{\nu}_{q}=-\frac{i}{2\sqrt{2}} \bigg ( q - \frac{\nu \sqrt{2} }{\pi} \bar{\Phi} \bigg )\psi^{\nu}_{q}, \textrm{ for } r \ge R\\ 
\frac{\partial}{\partial \theta}\psi^{\nu}_{q}=-\frac{i}{2\sqrt{2}} \bigg ( q + \frac{\nu \sqrt{2} }{\pi} \tilde{\Phi} \bigg )\psi^{\nu}_{q}, \textrm{ for } r \le R \\ \end{cases}
\mbox{,} \qquad q=  
\begin{cases}+1,\textrm{ Sub-Lattice A } ;\\ -1,\textrm{ Sub-Lattice B } ;\\ \end{cases}
\end{equation}

Which is valid, both for the static case and for the situation in which the wormhole has its matter content under rotation. 
Also, even when the lattice is subjected to the effect of gauge coupling (\(S_{sf}\)) related to the redshift term, the relationship mentioned above remains valid.

\section{Quantum Holonomy}\label{secIII}

In the model presented here, the behavior of electrons is governed by a massless Dirac equation. In molecular structures originating from the graphene matrix, quantum holonomies emerge as fundamental entities in understanding their topological and quantum aspects. Becoming indispensable tools for characterizing electronic properties on the lattice. Especially in graphene, characterized by Dirac cones at the corners of its Brillouin zone, leads to new physics in the nanoworld.
 
Moreover, the study of graphene wormholes has unveiled the potential for manipulating quantum information. Quantum gates and geometric phases acquired by fermions surrounding graphene derivatives open avenues for quantum computing applications \cite{Sjoqvist, DattaAdakScience}. In essence, quantum holonomies serve as a crucial theoretical framework and experimental probe, guiding the exploration of novel quantum phenomena and paving the way for groundbreaking advancements in graphene-based technologies.

For now, we will explore the potential to construct a quantum gate arising from the quantum phase acquired by fermions traveling through the surroundings of a graphene wormhole.
To achieve this, we will use a method for obtaining the geometrical phase ($\zeta(r,\bar{\Phi})$) of the system by applying the Weyl equation both outside and inside the throat (Eq. \ref{Weyl inside 2} and \ref{Weyl outside 2}). To accomplish this, we will employ the Dirac phase factor method, assuming the Dirac spinor is expressed as follows:

\begin{equation}
\Psi^{\nu}(t, r, \theta)=e^{\zeta (r,\bar{\Phi})}\Psi^{\nu}_{0}(t, r, \theta) = \exp{ \bigg ( - \int \Gamma_{\mu}(x)dx^{\mu} \bigg )}\Psi^{\nu}_{0}(t, r, \theta)
\end{equation}

Actually, the primary motivation for determining the spinorial connection arises from the realization that the partial derivative in the Weyl equation no longer guarantees gauge invariance within the Lorentz group \cite{KleinertGaugeFieldsinCondensedMatter}. Consequently, local Lorentz transformations must be incorporated into the fermion coupling. As a result, we can compute the spinorial connection by establishing a local reference frame at each point along the closed curve around the defect. Consequently, the holonomy matrix $U(r, \bar{\Phi}) = e^{\zeta(r,\bar{\Phi})}$ represents the parallel transport of a spinor along a path encircling both the inside and outside of the throat of the wormhole. The geometrical phase ($\zeta(r,\bar{\Phi})$) in both charts is given by:

\begin{equation}
\zeta(r,\bar{\Phi})_{in}= \frac{\ln (r)}{2\sqrt{2}\pi} \big [ \big ( v_{f}e^{\phi}+2i\omega_{\theta} \big ) \pi 
+ \nu \sqrt{2} \bar{\Phi} \sigma^{3} - \nu \sqrt{2} \bar{\Phi} \sigma^{1} - iv_{f}e^{\phi}\pi \sigma^{2}
\big ]
\mbox{,} \qquad \textrm{for} \qquad r \le R \mbox{,}  
\end{equation}
and
\begin{equation}
\zeta(r,\bar{\Phi})_{out}= \frac{\ln (r^{-2})}{4\sqrt{2}\pi} \big [ \big ( v_{f}e^{\phi}-2i\omega_{\theta} \big ) \pi
+ \nu \sqrt{2} \bar{\Phi} \sigma^{3} - \nu \sqrt{2} \bar{\Phi} \sigma^{1} - iv_{f}e^{\phi}\pi \sigma^{2}
\big ]
\mbox{,} \qquad \textrm{for} \qquad r \ge R \mbox{.}
\end{equation}
Where did we choose to write both chart in terms of effective flux out of throat ($\bar{\Phi}$). Further associating the Hausdorff formula
\begin{equation}
\exp(A)\exp(B)=\exp(A+B+{\frac {1}{2}}[A,B]+{\frac {1}{12}}([A,[A,B]]+[B,[B,A]])+...),
\end{equation}
with the matrices property $[\pi , \vec{\sigma}]=0$, we can found a useful version of the holonomic matrix:

\begin{equation}
U(r,\bar{\Phi})_{in}= ( \sqrt{r} )^{\delta}\exp \bigg ( \frac{\nu \bar{\Phi}}{2\pi}\ln(r)\sigma^{3} - \frac{\nu \bar{\Phi}}{2\pi}\ln(r)\sigma^{1} - \frac{iv_{f}e^{\phi}}{2\sqrt{2}}\ln(r)\sigma^{2} \bigg )
\mbox{,} \qquad \textrm{for} \qquad r \le R \mbox{,}
\end{equation}
and
\begin{equation}
U(r,\bar{\Phi})_{out}= \big ( \frac{1}{\sqrt{r}} \big )^{\delta^{*}}\exp \bigg ( -\frac{\nu \bar{\Phi}}{2\pi}\ln(r)\sigma^{3} + \frac{\nu \bar{\Phi}}{2\pi}\ln(r)\sigma^{1} + \frac{iv_{f}e^{\phi}}{2\sqrt{2}}\ln(r)\sigma^{2} \bigg )
\mbox{,} \qquad \textrm{for} \qquad r \ge R \mbox{.}
\end{equation} 
Where $\delta$ and $\delta^{*}$ are given by $\delta=\frac{v_{f}e^{\phi}+2i\omega_{\theta}}{\sqrt{2}}$ and $\delta^{*}=\frac{v_{f}e^{\phi}-2i\omega_{\theta}}{\sqrt{2}}$.
Furthermore, by expanding as $e^A \approx 1+A+\frac{A^2}{2!}+\dots$, and substituting $r\to \frac{r}{r_0}$, it is possible to express them as:

\begin{equation}
U(r,\bar{\Phi})_{in} \approx \big ( \frac{r}{r_{0}} \big )^{\frac{\delta}{2}}
\bigg ( 1 + \frac{\nu}{2}\ln \big ( \frac{r}{r_{0}} \big )\frac{\bar{\Phi}}{\pi} \sigma^{3} - \frac{\nu}{2}\ln \big ( \frac{r}{r_{0}} \big )\frac{\bar{\Phi}}{\pi}\sigma^{1} - \frac{iv_{f}e^{\phi}}{2\sqrt{2}}\ln \big ( \frac{r}{r_{0}} \big )\sigma^{2} \bigg )
\mbox{,} \qquad \textrm{for} \qquad r \le R \mbox{,} 
\end{equation}
and
\begin{equation}
U(r,\bar{\Phi})_{out} \approx \big ( \frac{r_{0}}{r} \big )^{\frac{\delta^{*}}{2}}
\bigg ( 1 - \frac{\nu}{2}\ln \big ( \frac{r}{r_{0}} \big )\frac{\bar{\Phi}}{\pi} \sigma^{3} + \frac{\nu}{2}\ln \big ( \frac{r}{r_{0}} \big )\frac{\bar{\Phi}}{\pi}\sigma^{1} + \frac{iv_{f}e^{\phi}}{2\sqrt{2}}\ln \big ( \frac{r}{r_{0}} \big )\sigma^{2} \bigg )
\mbox{,} \qquad \textrm{for} \qquad r \ge R \mbox{.}
\end{equation}

As it might be seen, running $r\to r_{0}$, the consequence is $U_{in}=U_{out}=1$. 
This condition ensures that the Hamiltonian in the fermion coupling has a unitary time evolution within the system. Consequently, the function $U(r, \bar{\Phi})$ remains regular everywhere. Additionally, a crucial consideration is the identification of $\bar{\Phi}$ as the principal control parameter. This not only elucidates the engineering underlying the system beyond the electronic density of states around the wormhole but also expands the spectrum of possibilities to manipulate future algoritm to quantum computation over graphene wormholes. 

\section{Summary and conclusion}\label{secIV}

This study introduces an effective field theory designed to capture the continuum limit of graphene wormholes within a redshift background, extending the groundwork established in prior works \cite{EvertonWormhole}. Additionally, we subject the material to rotation by doing \( \theta \to \theta^{'}=\theta +\omega_{\theta}t \). In such approach we focused on the idea of the redshift term as a potential interaction between the lattice and an external stretching field affecting the material. Described as proportional to the redshift term (\( \sim v_{f}e^{\phi}  \)). A new perspective has been provided here. Particularly relevant in scenarios involving materials under mechanical stress. In addition, our analysis reveals constraints on the azimuthal solutions (\( \psi_{q}^{\nu} (\theta) \)) of the Dirac equation for massless fermions. Furthermore, we also obtained quantum holonomies in this new scenario of the material under stress and rotation. We also observed that our results recovered the inertial case, and additionally, they extend to scenarios with no applied stretching effect. This work represents an additional effort to enhance our understanding of the interplay between non-inertial dynamics, redshift effects, and graphene physics, opening avenues for further exploration and practical applications in diverse contexts.

\section{Declarations}


{\bf Funding:} We thank CAPES , CNPQ and FAPESQ-PB for financial support.

{\bf Authors Contribution:} All authors have contributed equally to the manuscript.

{\bf Data Availability Statement:} Data sharing not applicable to this article as no datasets were generated or analyzed during the current study.

{\bf Conflict of Interests:} The authors declare no competing interests.

{\bf Ethical Conduct:} The authors declare that the research presented in this paper has been conducted with the highest standards of integrity and in accordance with relevant ethical guidelines.







\end{document}